\magnification=1200
\vsize=8.5truein
\hsize=6.3truein
\baselineskip=18truept
\parskip=4truept
\vskip 18pt
\def\today{\ifcase\month\or January\or February\or
March\or April\or May\or June\or July\or
August\or September\or October\or November\or
December\fi
\space\number\day, \number\year}
\centerline{{\bf On the power spectrum of undulations of simulated bilayers. }}
\centerline{by}
\centerline{J. Stecki }
\vskip 20pt
\centerline{Department III, Institute of Physical Chemistry,}
\centerline{ Polish Academy of Sciences, }
\centerline{ ul. Kasprzaka 44/52, 01-224 Warszawa, Poland}
\centerline{\today}
\vskip 40pt
\centerline {Abstract.}
\bigskip
The best  finite Fourier Series for a smooth surface $h(x,y)$ 
closest to the positions of heads of amphiphiles in the least-square sense,
agrees fully with the Fourier coefficients obtained by a direct summation
over raw data points. Both metods produce  structure factors $S(q)$ containing
all necessary features: small-q divergence, a minimum, the raise to the ubiquitous 
nearest neighbor peak near $q=2\pi/$(coll.diameter) and further peaks.
The Laurent series is also discussed.
\bigskip
\bigskip
The power spectrum of fluctuations {\it vel} structure factor $S(q)$ 
quantifies the undulations of a simulated bilayer. The 
deviations from planarity, of either monolayer or bilayer as a whole,
are described in terms of an imagined surface $h(x,y)$ represented as a finite
Fourier series $h(x,y)=\sum \tilde h({\bf q})\exp[i{\bf q}{\bf R}] $ .
The power spectrum is then $S=<\tilde h\tilde h^*>$ at any given $q$. 
At $x=|q^2|\to 0^+$ the theory[1-4] predicts
$ 1/S =kx^2+gx$ asymptotically, where $k$ is the
bending (or rigidity) coefficient[1] and $g$ is the fluctuation lateral tension 
 - slightly larger than the lateral tension $\Gamma$ determined from 
pressures[5].

The first point of this Letter is that the size of the region of $x$ where 
the asymptotic form is valid, can be estimated quantitatively by 
plotting the inverse $1/S$, further 
divided by $x$, as seen in Figure 14 in Reference[3]. It is extremely 
small[6,7], but as the quoted Figure[3] and many other such 
plots show, there is such a small range, it exists[8].   
A representation of $S(x)$ beyond that
minuscule asymptotic region would be very useful[6].
Because of the disparity of $g$ and
$\Gamma$ and for a variety of other reasons, almost all simulations 
choose the tensionless state ; there $S\sim 1/kx^2$[6,9-16].
Then, as already noted when the $q^{-4}$
dependence was uncovered[12] for the first time, an expansion in powers of $1/x$ 
works suprisingly well. 
I submit it is most understandable; $S$ originally had two poles 
$x_1=0, x_2=-g/k$  ( $x_2$  produces a new asymptote for $S$ in 
the floppy state where $g<0$[17,18]) merging into a double pole at
 $g=0$. Then $x^2S$ admits a Taylor expansion 
$x^2S=a_{-2} +a_{-1}x +a_0x^2 + a_1x^3 +... $ and the Laurent series is
$ S = a_{-2}/x^2 +a_{-1}/x +a_0 +a_1x +a_2x^2 +... .$
A preliminary fit of finite Laurent series is given in Supplementary Material,
making it clear that such representation is feasible[19]. 
The second point of this Note is then an extension of the original 
observation that power series of $x$ works very well but 
(1) there is no reason whatsoever to stop at two terms and 
(2) $a_n$ beyond $a_{-2}$ need not have direct meaning as proposed[6][8][12].

A most important feature of $S$ is the appearance of the 
nearest-neighbor (n.-n.) peak near 
$q=2\pi/\sigma$ ($\sigma$ being the collision diameter).
{\it This is a must as the bilayer is a liquid}. It seems most obvious 
that any interpretation of the $S(q)$ of the bilayer 
must include some bulk contribution of the liquid type.
The structure factors  and  the radial distribution functions $g(r)$ 
of bulk liquids are well known; 
$S_{bulk}$ raises from $S(0)$ (related to compressibility) to the n.-n. peak.
To that region of rising $S$ belongs all the liquid structure contained 
in $g(r)$. Correspondingly, $S$ 
of a bilayer  raises[17][9,10] above the asymptotic decay to  
{\it go through a minimum} and to raise to the
n.-n. peak. First such peak was shown some time ago[17] and in a 
recent work[9,10] several further peaks at still higher values of $q$ were 
obtained. 
It follows that any method which fails to 
show the  minimum in S(q) (and the n.n. peak if the $q$-range includes 
$2\pi /\sigma$), is therefore not to be trusted.

This caveat applies to methods based on gridding, which have never shown the 
ubiquitous nearest-neighbor peak. 

The structure of any liquid is described statistically as the density-density
correlation function $<\delta \rho (1)\delta \rho (2)>$, also 
for inhomogeneous fluids and for interfaces[20-22]. In our 
case of a nearly flat bilayer with 
periodic boundary conditions, it reduces to a function of $z_1,z_2,{\bf R}_{12}$
and when Fourier-transformed, to a function of  three variables, as
$H(z_1,z_2;q)$. We have determined $H$ for interfaces[22] and also
for coarse-grained (CG) bilayers[23]. Then, as it must, $H$ reproduces $S$ in a computation 
of 
$$ \int_0^L dz_1 \int_0^L dz_2 (z_1-z_0)H(z_1,z_2;q)(z_2-z_0). $$
The  method by which these well founded results are obtained uses the positions
of heads of amphiphilic molecules, $x_j,y_j,z_j$ and interprets $z_j$ as
the local "height" $h(x_j,y_j)$ so that 
$$ {\tilde h}({\bf q}) = const. \sum_j z_j\exp[i{\bf q R}_j]. $$
 The preselected set of values of ${\bf q}=(q_x,q_y)$ must 
follow the usual rules, but the above sums are approximations (in the spirit 
of the Monte-Carlo integration) because the theory of Fourier series defines
the coeficients such as $\tilde h$ as {\it integrals} over the function $h(x,y)$.

However, and that is the final point of this Letter,
 there exists a known modification of this last method, due to the
 Max-Planck(Potsdam) laboratory[24,25], little used[23-25],
which answers the following question:
" given a set of points $(z_j,x_j,y_j)$ what is the best surface $h(x,y)$ 
represented by a finite Fourier series?".
 The obvious "best" and most natural one  is 
"best in the least-squares sense".  Such $h(x,y)$  minimizes
the sum of squares of the deviations $h(x_j,y_j) - z_j$. The problem so posed
 is solved as a standard linear problem and the resulting finite Fourier 
series have been computed[23-25]. That is done for  each snapshot (each "time"
in the simulation run) and the resulting power spectra are then averaged over the
entire simulation run.
As $q_x=(2\pi/L_x)n_x$ and similarly $q_y$, we deal with $n_x,n_y$ .Once a 
maximum value of $n_y$ is assumed it defines a rectangle (half of a square)
and the set of resulting $q$-vectors is used. 
 The size of the matrix in  the linear problem is 81 for $n_{max}=4$,
 625 for 12, 1369 for 18,... . The limit is set by the number of data points, which 
reasonably ought to be no smaller than twice the number of coefficients, or 
larger. 

There can be no simpler nor more direct implementation of the "best" 
Fourier series $h(x,y)$ to represent the simulation data. And it supports 
the "sum-over-points" method, not the gridding method.
The tests,
applying the "sum-over-points" method and the "least-squares method"
to the same collection of positions of heads - demonstrate that these 
two methods of obtaining the power spectrum agree as well as one can 
expect or hope. 

Figure 1  shows such an example from one long run at an almost tensionless state
with $S$ produced by the "sum-over-points method", and three plots of $S$ by
the least-squares method. Clearly the agreement of shapes is excellent and
all show a nearest-neighbor peak or, if the range of $q$ is too small,
a minimum in $S(x)$.

It  may be noted as an aside that the low-q divergence does not change 
visibly with the decrease of $n_{max}$ even down to value of 4.

These tests demonstrate that the "sum-over-points method" produces correct and
reliable results.

\bigskip
\centerline {\bf References.}

\item {[1]}  W. Helfrich and R. M. Servuss, Nuovo Cimento 3D, 137 (1984);  
           see also  Helfrich, in {\it  Les Houches, Session XLVIII, 1988,
           Liquids at Interfaces} (Elsevier, New York, 1989).
         
\item {[2]} J.-B. Fourier, A. Adjari, and L. Peliti, Phys.Rev.Lett. 86,4970(2001).
\item {[3]} for a review of theory related to simulation (and some less common
            correlation functions), see J. Stecki,
            Advances in Chemical Physics 144, Chapter 3 (2010).
\item {[4]} J.F.Nagle and S.Tristam-Nagle, Biochem. et Biophys. Acta 1469,159-195(2000).
\item {[5]} J. Stecki, J. Chem. Phys. 120,3508(2004).
\item {[6]} Max C. Watson, Evgeni S. Penev, Paul M. Welch, and Frank L. H. Brown,
           J. Chem. Phys. 135,244701(2011)
\item {[7]} M. Deserno, see e.g. Macromol. Rapid.Commm.30,752(2009)
\item {[8]} E. R. May, A. Narang, and D. I. Kopelevich, Phys. Rev. E.76,021913(2007);
            the statement "which is the opposite of what is observed", is not correct.
\item {[9]} E.G.Brandt, A.R.Brown, J.N.Sachs, J.F.Nagle,and O.Edholm, Biophys.
            J. 100, 2104 (2011) and Supplement. 
\item {[10]} A.R.Braun,  E. G. Brandt, O. Edholm, J. F.Nagle, and J. N. Sachs, 
             Biophys. J. 100,2112(2011).
\item {[11]} E.Lindahl and O. Edholm, Biophys. J. 79,426(2000).
\item {[12]} R.Goetz, G. Gompper, and R. Lipowsky, Phys. Rev. Lett. 82,221(1999).
\item {[13]} O.Berger, O. Edholm, and F. Ja¨hnig, Biophys. J. 72,2002(1997).
\item {[14]} E.Lindahl and O. Edholm, J. Chem. Phys. 115,4938(2001).
\item {[15]} J.Wohlert and O. Edholm, J. Chem. Phys. 125,204703(2006).
\item {[16]} Q.Waheed  and O. Edholm,  Biophys. J. 97,2754(2009).
\item {[17]} J. Stecki, J. Chem. Phys. Rapid Comm. 122,111102 (2005).
\item {[18]} J. Stecki, cond-mat archive Dec.10,2004; J. Chem. Phys. 125,154902 (2006).
\item {[19]} Supplementary Material Document No. ...... for the Figure showing a rough 
             polynomial fit.
\item {[20]} R. Evans,  Adv.  Phys. 28, 143 (1979).
\item {[21]} P. Tarazona and R. Evans, Molec. Phys. 54,1357(1985)
\item {[22]} J.Stecki, J. Chem. Phys.  103,9763(1995), ibid. 108,3788(1998).
\item {[23]} J.Stecki, to be published (see also Cornell archive arxiv.org/cond-mat,2012). 
\item {[24]} A. Imparato, private communication of a remark by R.Lipowsky.
\item {[25]} A. Imparato, J. Chem. Phys. 124,154714(2006)(the area data).
\item {[26]} S.J.Marrink, H. J. Risselada, A. H. de Vries, J. Phys.Chem. B111,7812(2007).
\item {[27]} S.J.Marrinck, A.H.de Vries, and A.E.Mark, J.Phys.Chem. B 108,750(2003).
\item {[28]} R. Goetz and R. Lipowsky, J. Chem. Phys. 108,7397(1998).

\vfill\eject
\centerline {\bf Figure Captions.}
\bigskip

Caption to Fig.1
  
Comparison of $S(q)$ (by the direct-sum method) with that obtained by the least-squares
method (see text)  for 81, 625, and 441 Fourier coefficients. The identity 
of shapes demostrates the reliable accuracy of the "direct-sum-over-heads" 
 method of calculating $\tilde h({\bf q})$. A simplified CG model[26-28] of bilayer at 
$kT/\epsilon = 1.1$, overall density
$\rho=0.892$, monolayer with $N_1\sim 2000$ amphiphiles, box edge $L_x=38.25$, 5=4+1 
beads and Lennard-Jones solvent.

\vfill\eject

\centerline {\bf Supplementary Material.}
\bigskip
The Taylor series for $x^2S$ leads to a Laurent series for $S$, 
introduced in the main text as 
$$ S = a_{-2}/x^2 +a_{-1}/x +a_0 +a_1x +a_2x^2 +... .$$
with $x\equiv |q^2|$. 
It is of some interest to show that a polynomial obtained by cutting the
series at e.g. $a_4x^4$, is capable of reproducing reasonably well
the data points of $S(q)$ in the extensive range of $0<q<2\pi/\sigma$.
Here $\sigma$ is the collision diameter, common to all particles and beads  
in the Coarse Grained model used.

Figure 1S shows raw data for a tensionless state of such a coarse-grained 
model(CG) of a bilayer; the deviations from the fit of the Laurent series 
cut at $a_4$ are shown as vertical bars (impulses). 
Even without further refining
of this polynomial fit, the example shows that such representation is feasible. 
The same raw data obtained by the direct "sum-over-heads" method are compared 
in the main text with the obviously valid method based on least-squares 
approximation and introduced  by the Max-Planck (Potsdam) laboratory.

\centerline {\bf  Supplementary Material.   Figure 1S}
 
Caption to Fig.1S

The Figure 1S shows the so-called structure factor i.e fluctuation spectrum 
$S=<\tilde h \tilde h*>$ 
in the range about $q\in (0,2\pi/\sigma)$, plotted against $x=q^2$ and 
showing (1) the divergence; (2) the minimum; (3) the nearest-neighbor peak. 
Deviations from a simple fit to Laurent series with 7 terms, are shown 
with vertical lines (impulses). A simplified CG model of bilayer at 
$kT/\epsilon = 1.1$, overall density
$\rho=0.892$, monolayer with $N_1\sim 2000$ amphiphiles, box edge $L_x=38.25$,
area $L_x^2$, 6.E6 timesteps, 5=4+1 beads and Lennard-Jones solvent.

\vfill\eject\end
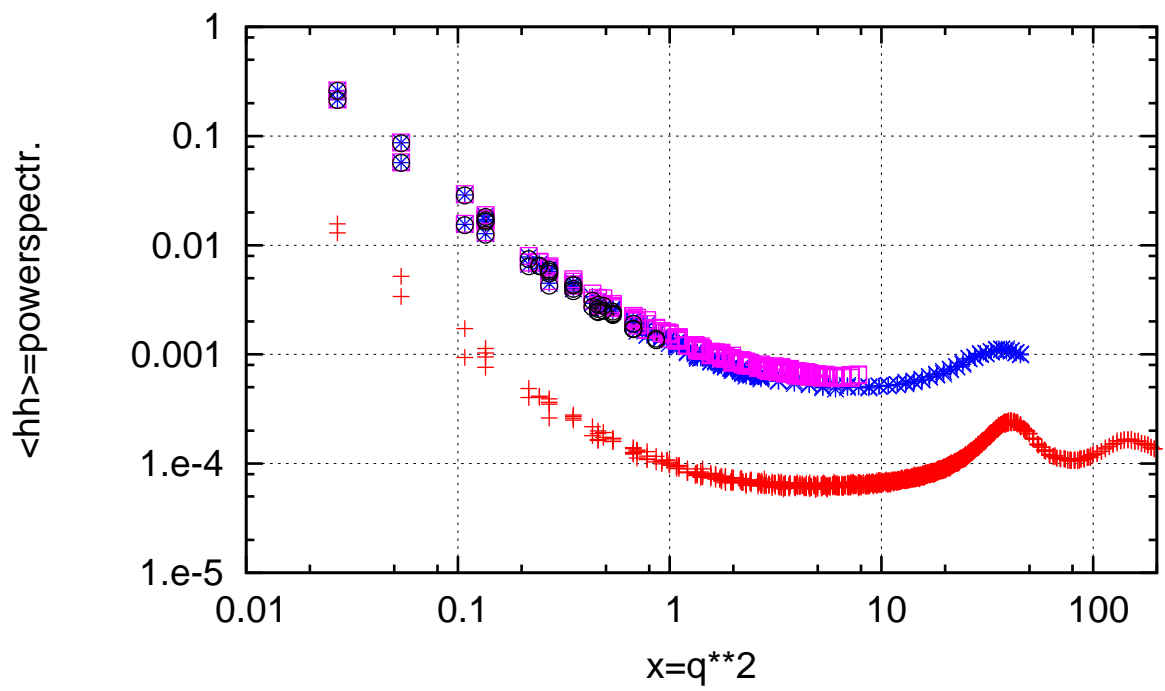
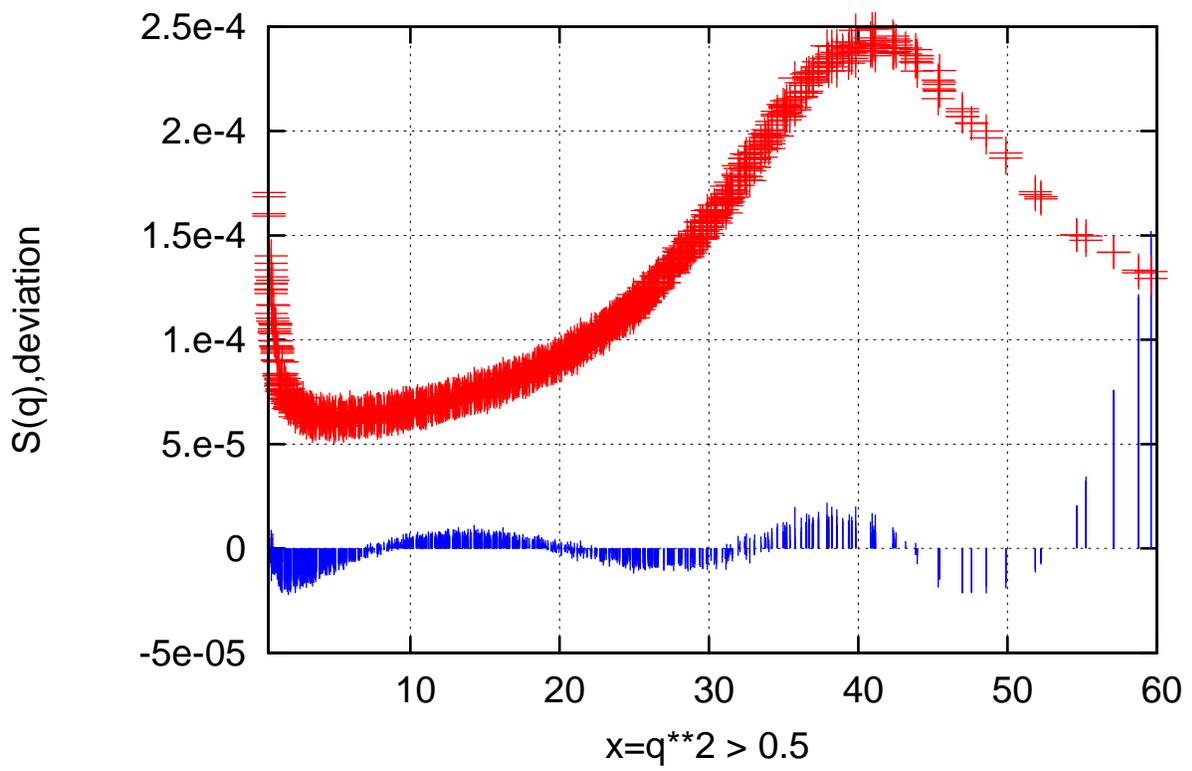
\bye